\documentclass{article}
\usepackage{icml2003}
\include{epsf}
\usepackage{mlapa}

\usepackage{times}
\usepackage{verbatim}
\usepackage{amssymb}
\usepackage{amsmath}
\usepackage{amsfonts}

\newtheorem{thm}{Theorem}
\newtheorem{defin}{Definition}
\newtheorem{game_description}{Game family description}
\newtheorem{lemma}{Lemma}

\def\QuadSpace{\vspace{0.25\baselineskip}}
\def\HalfSpace{\vspace{0.5\baselineskip}}

\def\EndProof{ \quad \vrule width 1ex height 1ex depth 0pt \newline }

\newenvironment{proof}{\QuadSpace\par\noindent{\bf Proof}:}{\EndProof\HalfSpace \vspace{-0.15in}}

\font\sf=cmss10
\newcommand{\Nats}{{\hbox{\sf I\kern-.13em\hbox{N}}}}   
\newcommand{\Reals}{{\hbox{\sf I\kern-.14em\hbox{R}}}}  
\newcommand{\Ints}{{\hbox{\sf Z\kern-.43emZ}}}          
\newcommand{\CC}{{\hbox{\sf C\kern -.48emC}}}           
\newcommand{\QQ}{{\hbox{\sf C\kern -.48emQ}}}           

\begin{document}

\twocolumn[ \icmltitle{BL-WoLF: A Framework For Loss-Bounded Learnability In Zero-Sum Games}

\icmlauthor{Vincent Conitzer}{conitzer@cs.cmu.edu}
\icmladdress{Computer Science Department,
Carnegie Mellon University,
5000 Forbes Avenue, 
Pittsburgh, PA 15213}
\icmlauthor{Tuomas Sandholm }{sandholm@cs.cmu.edu}
\icmladdress{Computer Science Department, Carnegie Mellon University, 5000 Forbes Avenue, 
Pittsburgh, PA 15213}
\vskip 0.3in
\vskip 0.3in
]

\begin{abstract}
We present BL-WoLF, a framework for learnability in repeated zero-sum
games where the cost of learning is measured by the losses the
learning agent accrues (rather than the number of rounds). 
%
%
The game is adversarially chosen from some family that the learner
knows. The opponent knows the game and the learner's learning
strategy.
%
%
The learner tries to either not accrue losses, or to quickly learn
about the game so as to avoid future losses (this is consistent with
the Win or Learn Fast (WoLF) principle; BL stands for ``bounded loss'').  Our framework allows
for both probabilistic and approximate learning. The resultant notion of {\em BL-WoLF}-learnability can
be applied to any class of games, and allows us to measure the
inherent disadvantage to a player that does not know which game in the
class it is in.

%

We present {\em guaranteed BL-WoLF-learnability} results for families
of games with deterministic payoffs and families of games with
stochastic payoffs.  We demonstrate that these families are {\em
guaranteed approximately BL-WoLF-learnable} with lower cost.  We then
demonstrate families of games (both stochastic and deterministic) that
are not guaranteed BL-WoLF-learnable.  We show that those families,
nevertheless, are {\em BL-WoLF-learnable}.  To prove these results,
we use a key lemma which we derive.\footnote{This material is based upon work supported by the National Science
Foundation under CAREER Award IRI-9703122, Grant IIS-9800994, ITR
IIS-0081246, and ITR IIS-0121678.}
\end{abstract}

\section{Introduction}

When an agent is inserted into an unfamiliar environment with some
objective, two goals present themselves. The first is to learn the
relevant aspects of the environment, so that eventually, its behavior
is optimal or near optimal with regard to the given objective. The
second is to minimize the cost of learning to behave well. This can be
done by minimizing the time necessary to learn enough to perform well,
but also by ensuring that its behavior in the learning process, while
not yet optimal or near optimal, is at least reasonably good with
regard to the objective. There is often an exploration/exploitation tradeoff here: attempting to
learn fast often requires disastrous short term results, while slow
learning may accumulate large losses even if the loss per unit time is
small.

Learning in games (for a review, see~\cite{Fudenberg98:Theory}) is
made additionally difficult because the learner is confronted with
another player (or multiple other players). If the other player plays
in a predictable, repetitive manner, this is no different from
learning in an impersonal, disinterested environment. Usually, however, the
other player changes its strategy over time. One reason for this may
be that the other player is also learning. A less benign reason, however, may be that the opponent is
aware of the learner's predicament and is trying to exploit its
superior knowledge. This is the case that we study.

In the case where an opponent is trying to exploit the learner's lack
of knowledge about the game, it becomes especially important to focus
on the cumulative cost of learning rather than the time the learning
takes. It is likely that the opponent will allow the learner to learn
the game very quickly, if the opponent can take tremendous advantage
of the learner in the short run. A learning strategy on the learner's part that
allows this should not be considered good. On the other hand, a
learning strategy that may learn the relevant structure of the game
only very late or even never at all, but allows the opponent to take
only minimal advantage, should be considered good. This analysis is
consistent with numerous learning results in the game theory and
machine learning literatures which guarantee convergence to a strategy
OR that the payoffs approach those of the
equilibrium (e.g.~\cite{Jehiel01:Learning,Singh00:Nash}). It suggests a
Win-or-Learn-Fast, or WoLF, approach (a term coined by Bowling and
Veloso~\cite{Bowling02:Multiagent}, though they actually just pursued
convergence results). Various previous work has considered the case where
learning players are concerned with their long-term losses, for
instance when players have beliefs about the opponents'
strategies~\cite{Kalai93:Rational}. 

Much of the prior work on learning in games in the machine learning
literature did not consider such a metric of the performance of a
learning strategy~\cite{Littman94:Markov,Hu98:Multiagent}. In
contrast, our work is especially closely related to recent work by
Brafman and Tennenholtz on learning in stochastic
games~\cite{Brafman00:Near}, where the opponent can make it difficult
to learn parts of the game, leading to a complex exploration
vs. exploitation tradeoff (building upon closely related
work~\cite{Kearns98:Near,Monderer97:Dynamic}); and on learning
equilibrium~\cite{Brafman02:Efficient}, where the agents' learning
algorithms over a class of games are considered as strategies
themselves.

In another strand of research, Auer {\em et al.} also study the
problem of learning a game with the goal of minimizing the cumulative
loss due to the learning process, with an adversarial
opponent~\cite{Auer95:Gambling}. (This problem is studied towards the
end of that paper.) They study the case where the learner knows
nothing at all about the game (except the learner's own actions and
bounds on the payoffs), and they derive an algorithm for this general case,
which improves over previous algorithms by Ba\~{n}os~\cite{Banos68:Pseudo}
and Megiddo~\cite{Megiddo80:Repeated}. (Some closely related research
makes the additional assumption that the learner, at the end of each
round, gets to see the expected payoff for {\em all} the actions the
learner might have chosen, given the opponent's mixed
strategy~\cite{Freund99:Adaptive,Fudenberg95:Consistency,Foster93:Randomization,Hannan57:Approximation}. 
We will not make this
assumption here.) The main difference between that line of work and
the framework presented here is that our framework allows the learner
to take advantage of partial knowledge about the game (that is,
knowledge that the game belongs to a certain family of games). This
allows the learner to potentially perform much better than a
general-purpose learning algorithm.\footnote{The gap between the two
approaches in the case of partial knowledge of the game may be
partially bridged through the use of different {\em experts}~\cite{Cesa-Bianchi97:How}, who make
recommendations to the agents as to which actions to play. For
instance, the Auer {\em et al.} paper~\cite{Auer95:Gambling} also
studies how to learn which is the best of a set of given
experts. These experts could capture some of the known structure of
the game: for instance, there could be an expert recommending the
optimal strategy for each game in the family. However, the learning
algorithm for deciding on an expert will typically still not make full
use of the known structure.}

In this paper, we introduce the {\em BL-WoLF} framework, where a
learner's strategy is evaluated by the loss it can expect to accrue as
a result of its lack of knowledge. (We consider the worst-case loss
across all possible opponents as well as all possible games within the
class considered. BL stands for ``bounded loss''.)  We present a {\em guaranteed} version of learnability
where the learner is guaranteed to lose no more than a given amount,
and a {\em nonguaranteed} version where the agent loses no more than a
given amount {\em in expectation}. We also allow for {\em approximate}
learning in both cases, where we only require that the agent comes
close to acting optimally.
The framework is applicable to
any class of (repeated) games, and allows us to measure the inherent
disadvantage in that class to a player that initially cannot
distinguish which game is being played. It does not assume a
probability distribution over the games in the class.

We do not consider difficulties of computation in games; rather we
assume the players can deduce all that can be deduced from the
knowledge available to them. While some of the most fundamental
strategic computations in game theory have
unknown~\cite{Papadimitriou01:Algorithms} or
high~\cite{Conitzer02:Nash} complexity in general,
zero-sum~\cite{Luce57:Games} and repeated~\cite{Littman03:Polynomial}
games tend to suffer fewer such problems, thereby at least partially
justifying this approach. In the game families in this paper, computation will be simple.

The rest of this paper is organized as follows. In
Section~\ref{se:definitions}, we give some basic definitions and known
results. We present guaranteed BL-WoLF learnability in
Section~\ref{se:g}, and its approximate version in
Section~\ref{se:ga}.  We present nonguaranteed BL-WoLF learnability
in Section~\ref{se:n}, and its approximate version in
Section~\ref{se:na}.

\section{Basic definitions}
\label{se:definitions}

Throughout the paper, there will be two players: the learner (player
$1$) and the opponent (player $2$). Because we try to assess the
worst-case scenario for the learner, restricting ourselves to only one
opponent is without loss of generality---if there were multiple
opponents, the worst-case scenario for the agent would be when the
opponents all colluded and acted as a single opponent.

In this paper, the two players play a one-shot (or {\em stage})
zero-sum game over and over. Player $2$ knows the game; player $1$ (at
least initially) only knows that it is in a larger {\em family} of
games.  In this section, we will first define the stage game, and
discuss what it means to play it well on its own. We then define the
uncertainty that player $1$ has about the game. Finally, we define
what strategies the players can have in the repeated game. Definitions
on what it means for the learner to play the repeated game well are
presented in later sections.

\subsection{Zero-sum game theory for the stage game}

\begin{defin}
A {\em (stage) game} consists of sets of actions $A_1, A_2$ for players $1$
and $2$ respectively, together with (in the case of deterministic payoffs) a function $u: A_1 \times A_2
\rightarrow U_1 \times U_2$, where $U_i$ is the space of possible
utilities for player $i$ (usually simply $\Reals$); or (in the case of stochastic payoffs) a function $p_u: A_1 \times A_2
\rightarrow \mathcal{P}(U_1 \times U_2)$, where $\mathcal{P}(U_1
\times U_2)$ is the set of probability distributions over utility
pairs. We say the game is {\em zero-sum} if the utilities of agent $1$ and
$2$ always sum to a constant.
\end{defin}

We often say that the random selection of an outcome in a game with
stochastic payoffs is done by {\em Nature}. For the following
strategic aspects, it is irrelevant whether Nature plays a part or
not.


\begin{defin}
A {\em (stage-game) strategy} for player $i$ is a probability
distribution over $A_i$. (If all of the probability mass is on one
action, it is a {\em pure strategy}, otherwise it is a {\em mixed
strategy}.) A pair of strategies $\sigma_1, \sigma_2$ for players $1$
and $2$ are in {\em Nash equilibrium} if neither player can obtain
higher expected utility by switching to a different strategy, given
the other player's strategy. A strategy $\sigma_i$ is a {\em maximin}
strategy if $\sigma_i \in \arg \max_{\sigma_i} \min_{\sigma_{-i}}
E[u_i|\sigma_i,\sigma_{-i}]$.\footnote{Here we use the common game theory
notation $-i$ for ``the player other than $i$''.}
\end{defin}

The following theorem shows the relationship between maximin
strategies and Nash equilibria in zero-sum games. Informally, it shows
why, against a knowledgeable opponent, a player is playing well if and only
if that player is playing a maximin strategy.

\begin{thm}[Known]
In zero-sum games, a pair of strategies $\sigma_1, \sigma_2$
constitute a Nash equilibrium if and only if they are both maximin
strategies. The expected utility that each player gets in an
equilibrium is the same for every equilibrium; this expected utility
(for player $1$) is called the {\em value} $V$ of the game.
\end{thm}

Thus, player $1$ is guaranteed to get an expected utility of at least $V$ by
playing a maximin strategy (and player $2$ can make sure player $1$
gets at most $V$ by playing a maximin strategy). We call a strategy
$\sigma_1$ an {\em $\epsilon$-approximate} maximin strategy if it
guarantees an expected utility of $V-\epsilon$. The {\em stage-game
loss} of player $1$ in playing the stage game once is $V$ minus the
utility player $1$ received.

\subsection{What player $1$ does not know}

Player $1$ (at least initially) does not know which of a {\em family}
of zero-sum stage games is being played. Such a family is defined as
follows:

\begin{defin}
A {\em parameterized family of stage games with deterministic (stochastic)
payoffs} is defined by action sets $A_1$ and $A_2$, a parameter space
$K$, and a function $g:K \rightarrow {\mathcal G_d}(A_1,A_2)$ ($g:K \rightarrow {\mathcal G_s}(A_1,A_2)$), where
${\mathcal G_d}(A_1,A_2)$ (${\mathcal G_s}(A_1,A_2)$) is the set of all zero-sum stage games
with deterministic (stochastic) payoffs with action sets $A_1, A_2$.
%
\end{defin}

Here, player $1$ does not know the parameter $k \in K$ corresponding
to the game being played.\footnote{The parameter space $K$ is not strictly
necessary (all that matters for our purposes is the subset of games in
the image of $g$), but it is often convenient to think of the missing
knowledge as a parameter of the game.} In the examples in this paper,
the elements of $K$ will take many forms, such as integers,
permutations, and subsets. Player $1$ can eliminate values of $K$ on
the basis of outcomes of games played.

We note that there is no probability distribution on the family of
games. Rather, we assume the game is adversarially chosen relative to
the learner's learning strategy.



\subsection{Strategies in the repeated game}

A strategy in the repeated game (in the case of player $1$, a {\em
learning} strategy) prescribes a stage-game strategy given any history
of what happened in previous stage games. Thus, the stage-game
strategy can be conditional on the players own past actions, the other
player's past actions, and past payoffs. In our paper, it will usually
be sufficient for it to just be conditional on player $1$'s knowledge
about the game. To evaluate how well player $1$ is doing, we define
player $1$'s {\em (cumulative) loss} as the sum of all stage-game
losses. Thus, if player $1$ knew the game, playing the maximin
strategy forever would give an expected loss of at most $0$ against
any opponent. (We do not use a discounting rate; rather, when we
aggregate utilities, we consider the sum of utilities across finite
numbers of games.)\footnote{It is crucial to distinguish between the learning strategy and the
stage-game strategies it produces. When we talk about a maximin
strategy or about learning a strategy, we are referring to stage-game
strategies. Otherwise, we will make it clear which one we refer to.}

\section{Guaranteed BL-WoLF-learnability}
\label{se:g}

In the simplest form of learning in our framework, there is a learning strategy
for player $1$ such that, having accumulated a given amount of loss,
player $1$ is {\em guaranteed} to know enough about the game to play
it well. In this section, we give the formal definition of this type
of learnability, and demonstrate that some example game families
(including games with stochastic payoffs) are learnable in this sense.


\begin{defin}
A parameterized family of games is {\em guaranteed BL-WoLF-learnable}
with loss $l$ if there exists a learning strategy for player $1$ such that, for
any game in the family, against any opponent, the loss incurred by
player $1$ before learning enough about the game to construct a
maximin strategy is never more than $l$.
\end{defin}


\begin{game_description}\footnote{When describing a family of games, we usually describe the family for
some arbitrary variables. Thus, the definition starts with ``For given
$X$, the family of games $Y$ is defined by...'' These $X$ are {\em
not} the parameters to be learned; they are known by
everyone. Effectively, we have a family of families of games, one
family for each value of $X$. The parameter $k \in K$ to be learned
with such a family is pointed out in the end of the definition, under
the header {\em Player $1$ initially does not know:}.}
For a given $n$, the game family {\em get-close-to-the-target} is
defined as follows. Players 1 and 2 both have action space
$A=\{1,2,\ldots,n\}$. The outcome function is defined by a parameter $k \in
\{1,2,\ldots,n\}$, that the players try to get close to. Given the
actions by the players, the outcome of the game is as follows (winning gives utility $1$, losing utility $-1$):

$\bullet$ If $|a_i - k| < |a_{-i} - k|$, then player $i$ wins;

$\bullet$ If $a_1 = a_2 = a \neq k$, player $1$ wins if $a < k$, and
player $2$ wins if $a > k$;

$\bullet$ Otherwise ($a_1 - k = k - a_2$), we have a
draw.

{\em Player $1$ initially does not know:} the parameter
$k$.
\end{game_description}

\begin{thm}
The game family {\em get-close-to-the-target} is guaranteed BL-WoLF-learnable with loss $\lceil \log(n) \rceil$.
\label{th:get}
\end{thm}
\vspace{-.12in}
\begin{proof}
We first observe that if we ever have a draw, player $1$ can
immediately infer $k$---it is the average of the players'
actions. Also, after any number of rounds, the set of possible values
for $k$ that are consistent with the outcomes so far is always an
interval $\{k^{min},k^{min}+1,\ldots,k^{max}\}$. (The set of possible
values for $k$ that are consistent with a single outcome is always an
interval, and the intersection of two intervals is always an
interval.) Now consider the following learning strategy for player $1$: always
play the action in the middle of the remaining interval, $a_1 =
\lfloor \frac{k^{min}+k^{max}}{2} \rfloor$. If player $1$ loses, it
can be concluded that $k$ is on the side of $a_1$ where player $2$
played. ($a_2 \leq a_1 \Rightarrow k < a_1$ and $a_2 > a_1 \Rightarrow k
> a_1$.) Thus the remaining interval is cut in half (sometimes the
remainder is less than half, because the action player $1$ played is
also eliminated; it is never more). So, after $\lceil \log(n) \rceil$
losses, player $1$ knows $k$, and the maximin strategy
(which is simply to play $k$).
\end{proof}

The parameter to be learned need not always be an
integer. In the next example, it is a permutation of a finite
set.

\begin{game_description}
For given $m > 2$ and $n$, the game family {\em
generalized-rock-paper-scissors-with-duds} is defined as
follows. Players 1 and 2 both have action space
$A=\{1,2,\ldots,m+n\}$. The outcome function is defined by a
permutation $f:\{1,2,\ldots,m+n\} \rightarrow \{1,2,\ldots,m+n\}$. The
set of {\em duds} is given by $\{i: m+1 \leq f(i) \leq m+n\}$. Given the
actions by the players, the outcome of the game is as follows (winning gives utility $1$, losing utility $-1$):

$\bullet$ If only one player plays a dud, that player loses;

$\bullet$ If neither player plays a dud and $f(a_i)=f(a_{-i})+1 (mod m)$,
player $i$ wins (effectively, the nonduds are arranged in a circle,
and playing the action right after your opponent's in the circle gives
you the win);

$\bullet$ Otherwise, we have a draw.

{\em Player $1$ initially does not know:} the permutation $f$. 
(We observe that for $m=3$ and $n=0$, we have the classic
rock-paper-scissors game.)
\end{game_description}

\begin{thm}
The game family {\em generalized-rock-paper-scissors-with-duds} is
guaranteed BL-WoLF-learnable with loss $m-1$ if $m$ is even, or with
loss $m$ if $m$ is odd. If $n=0$, it is guaranteed BL-WoLF-learnable
with loss $0$.
\label{th:rps}
\end{thm}
\vspace{-.12in}
\begin{proof}
Consider the following learning strategy for player $1$. Keep playing action
$1$ first; then, whenever player $2$ wins a round, switch to the
action that he just won with, and keep playing that until player $2$
wins again. Because it is impossible to win when playing with a dud,
the first action that player $2$ wins a round with must be a
nondud. After this, player $2$ can win only by playing the next action
in the circle of nonduds. Thus, every loss reveals the next element in
the circle. Thus, after $m$ losses, the whole circle of nonduds is
revealed and player $1$ can choose a maximin strategy. (For instance,
randomizing uniformly over the nonduds.) In the case where $m$ is
even, only $m-1$ losses are needed, as this reveals the whole circle
but one---and when $m$ is even, it is a maximin strategy to randomize
uniformly over all the nonduds $i$ such that $f(i)$ is even (or all
the nonduds $i$ such that $f(i)$ is odd),
and we can determine one of these two sets even with a ``gap'' in the
circle. Finally, if $n=0$, we need not learn anything about $f$ at all: simply randomize uniformly
over all the actions.
\end{proof}

Game families with stochastic payoffs can also be guaranteed
BL-WoLF-learnable. The following modification of the previous game
illustrates this.

\begin{game_description}
The game family {\em
random-orientation-generalized-rock-paper-scissors-with-duds} is
defined exactly as {\em generalized-rock-paper-scissors-with-duds},
except each round, Nature flips a coin over the orientation of the
circle of nonduds. That is, with probability $\frac{1}{2}$, if neither
player plays a dud and $f(a_i)=f(a_{-i})+1 (mod m)$, player $i$ wins;
otherwise, if neither player plays a dud and $f(a_i)=f(a_{-i})-1 (mod
m)$, player $i$ wins. The other cases are as before: nonduds still
(always) beat duds, and we have a draw in any other case.

{\em Player $1$ initially does not know:} the permutation $f$.
\end{game_description}

\begin{thm}
The game family {\em random-orientation-generalized-rock-paper-scissors-with-duds} is
guaranteed BL-WoLF-learnable with loss $1$ (or loss $0$ if $n=0$).
\end{thm}
\vspace{-.12in}
\begin{proof}
We simply observe that playing {\em any} nondud action is a maximin
strategy in this case. (Any nondud action is as likely to lose against
it as to win.) Player $1$ will know such an action upon being beaten
once (or, if there are no duds, player $1$ will know such an action
immediately).
\end{proof}


\section{Guaranteed approximate BL-WoLF-learnability}
\label{se:ga}

We now introduce approximate BL-WoLF-learnability.

\begin{defin}
A parameterized family of games is {\em guaranteed approximately
BL-WoLF-learnable} with loss $l$ and precision $\epsilon$ if there
exists a learning strategy for player $1$ such that, for any game in the
family, against any opponent, the loss incurred by player $1$
before learning enough about the game to construct an
$\epsilon$-approximate maximin strategy is never more than $l$.
\end{defin}

To save space, we only present one straightforward approximate
learning result on a game family we have studied already, to
illustrate the technique. A similar result can be shown for {\em generalized-rock-paper-scissors-with-duds}.

\begin{thm}
The game family {\em get-close-to-the-target} is guaranteed
approximately BL-WoLF-learnable with loss $r$ and precision
$1 - \frac{2^{r}}{n}$ (for $r < \log(n)$).
\end{thm}
\vspace{-.12in}
\begin{proof}
We consider the same learning strategy as before, where we always play the
middle of the remaining interval. After $r$ losses, the remaining
interval has size at most $\frac{n}{2^r}$. Randomizing over the
remaining interval will give at least a draw with probability at least
$\frac{1}{\frac{n}{2^r}}= \frac{2^r}{n}$.
\end{proof}

\section{Nonguaranteed BL-WoLF-learnability}
\label{se:n}

Guaranteed learning (even approximate) is not always possible. In many
games, no matter what learning strategy player $1$ follows, it is possible that
an unlucky sequence of events leads to a tremendous loss for player
$1$ without teaching player $1$ anything about the game. Such unlucky
sequences of events can easily occur in games with stochastic payoffs, but
also in games with deterministic payoffs where player $1$'s only hope of
learning against an adversarial opponent is by using a mixed strategy. (We will
see examples of both these cases later in this section.) Nevertheless,
it is possible that there are learning strategies in these games that {\em in
all likelihood} will allow player $1$ to learn about the game without
incurring too much of a loss. In this section, we present a more
probabilistic definition of learnability; we show that it is strictly
weaker than guaranteed BL-WoLF-learnability; we present a useful
lemma for showing this type of BL-WoLF-learnability; and we apply
this lemma to show BL-WoLF-learnability for some games that are not
guaranteed BL-WoLF-learnable.

\subsection{Definition}

\begin{defin}
A parameterized family of games is {\em BL-WoLF-learnable} with loss
$l$ if there exists a learning strategy for player $1$ such that, for any game
in the family, against any opponent, and for any integer $N$,
player $1$'s expected loss over the first $N$ rounds is at most $l$.
\end{defin}

We now show that BL-WoLF-learnability is indeed a weaker notion than
guaranteed BL-WoLF-learnability.

\begin{thm}
If a parameterized family of games is guaranteed BL-WoLF-learnable
with loss $l$, it is also BL-WoLF-learnable with loss $l$.
\end{thm}
\vspace{-.12in}
\begin{proof}
Given the learning strategy $\sigma$ that will allow player $1$ to learn enough
about the game to construct a maximin strategy with loss at most $l$,
consider the learning strategy $\sigma'$ which plays $\sigma$ until the maximin
strategy has been learned, and plays the maximin strategy forever
after that. Then, after $N$ rounds, if we are given that no maximin
strategy has been learned yet, the loss must be less than $l$. Given
that a maximin strategy was learned after $i \leq N$ rounds, the loss
up to and including the $i$th round must have been less than $l$, and
the expected loss after round $i$ is at most $0$ (because a maximin strategy
was played in every round after this). It follows that the expected
loss is at most $l$.
\end{proof}

\subsection{A central lemma}

The next lemma will help us prove the
BL-WoLF-learnability of games that are not guaranteed
BL-WoLF-learnable.

\begin{lemma}
Consider a learning strategy for player $1$ that plays the same stage-game
strategy every round until some learning event. (Call a sequence of
rounds between learning events throughout which the same stage-game
strategy is played an {\em epoch}.) Suppose that the following two
facts hold for any game in the parameterized family:

$\bullet$ For any epoch $i$'s stage-game strategy $\sigma_1^i$ for
player $1$, any stage-game strategy $\sigma_2$ for player $2$ will
either with nonzero probability cause the learning event that changes
the epoch to $i+1$, or will not give player $2$ any advantage
(i.e. player $1$'s expected loss from the round when player $2$ plays
$\sigma_2$ is at most $0$).

$\bullet$ For any of those strategies $\sigma_2$ that with nonzero
probability cause the learning event that changes the epoch to $i+1$,
we have $\frac{\lambda(\sigma_1^i,\sigma_2)}{p^i(\sigma_1^i,\sigma_2)} \leq
c_i$ for some given $c_i \geq 0$. (Here $\lambda(\sigma_1^i,\sigma_2)$ is the
expected one-round loss to player $1$, and $p^i(\sigma_1^i,\sigma_2)$
is the probability of this round causing the learning event that
changes the epoch to $i+1$.)

Then with this learning strategy, the family of games is BL-WoLF-learnable with
loss $\sum\limits_{i}c_i$.
\label{le}
\end{lemma}
\vspace{-.12in}
\begin{proof}
Given the number $N$ of rounds, divide up player $1$'s total loss $l$
over the epochs. That is, for epoch $i$, we have $l_i = \sum\limits_{j
\leq N, j \in i}\lambda_j$ where $\lambda_j$ is player $1$'s loss in
round $j$; and $l=\sum\limits_{i}l_i$. Consider now an opponent that
seeks to maximize the expectation of a given $l_i$.  If there is no
action that gives this opponent any advantage in this epoch (player
$1$ is already playing a maximin strategy), the expected value of
$l_i$ cannot exceed $0 \leq c_i$. If there is an action that gives the
opponent some advantage, by the first fact, it causes the end of the
epoch with some nonzero probability. In this case, playing an action
that does not cause the end of the epoch with some nonzero probability
is a bad idea for the opponent, because doing so gives the opponent no
advantage and just brings us closer to the limit to the number of
rounds $N$. So we can presume that the opponent only plays actions
that cause the end of the epoch with some nonzero probability. Now
suppose that there is no limit to the number of rounds, but the
opponent is still restricted to playing actions that cause the end of
the epoch with some nonzero probability. (This is still a preferable
scenario to the opponent.) In this scenario, we have
$\max_{\sigma_2}(E[l_i])=\max_{\sigma_2}(\lambda(\sigma_1^i,\sigma_2)+(1-p^i(\sigma_1^i,\sigma_2))\max_{\sigma_2}(E[l_i]))$,
and it follows that
$\max_{\sigma_2}(E[l_i])=\max_{\sigma_2}(\frac{l(\sigma_1^i,\sigma_2)}{p^i(\sigma_1^i,\sigma_2)})
\leq c_i$. It follows that the expectation of any $l_i$ is bounded by
$c_i$, for any opponent. Thus (by linearity of expectation) the total
expected loss is bounded by $\sum\limits_{i}c_i$.
\end{proof}


\subsection{Specific game families}

We first give an example of a game family with stochastic payoffs
where guaranteed BL-WoLF learning is impossible because Nature might
be noncooperative.

\begin{game_description}
For given $n, p_1, p_2, r_1, r_2$, the game family {\em get-close-to-one-of-two-targets} is defined exactly as
{\em get-close-to-the-target}, except now there are two $k_1,k_2 \in
\{1,2,\ldots,n\}$, with $k_1 \neq k_2$. Each round, Nature randomly
chooses which of the two is ``active'' ($k_j$ is active with
probability $p_j$). The winner is the player that would
have won {\em get-close-to-the-target} with that $k_j$. The utility of
winning is dependent on $j$: the winner receives $r_j$ (with $r_1 \neq r_2$; the loser gets $0$).

{\em Player $1$ initially does not know:} the parameters
$k_1$ and $k_2$. 
\end{game_description}


{\em Get-close-to-one-of-two-targets} is not guaranteed
BL-WoLF-learnable, for the following reason. Consider the scenario
where $k_1$ is to the left of the middle, $k_2$ is to the right of the
middle, and player $2$ is consistently playing exactly in the
middle. Now, regardless of which action player $1$ plays, for one of
the $k_i$, player $2$ will win if this $k_i$ is active; and player $1$
will be able to infer nothing more than which side of the middle that
$k_i$ is on. Thus, if Nature happens to keep picking $k_i$ in this manner, player $1$ will accumulate a huge loss without
learning anything more than which sides of the middle the $k_i$ are
on. It is easy to show that, if one of the $k_i$ is much more likely
and valuable than the other, this can leave us arbitrarily far away
from knowing a maximin strategy.  Nevertheless, with the probabilistic
definition, {\em get-close-to-one-of-two-targets} is BL-WoLF-learnable
for a large class of values of the parameters $p_1$, $p_2$, $r_1$, and
$r_2$ (which includes those cases where one of the $k_i$ is much more
likely and valuable than the other), as the next theorem shows.

\begin{thm}
If $p_1r_1 \geq 2p_2r_2$, then the game family {\em
get-close-to-one-of-two-targets} is BL-WoLF-learnable with loss
$\lceil \log(n) \rceil r_1$.
\end{thm}
\vspace{-.12in}
\begin{proof}
First we observe that if $p_1r_1 \geq 2p_2r_2$, then playing $k_1$ is
then a maximin strategy. (To prove this, all we need to show is that
both players playing $k_1$ is an equilibrium. When the other player is
playing $k_1$, also playing $k_1$ gives expected utility at least
$\frac{p_1r_1}{2}$, and any other pure strategy gives at most
$p_2r_2$, which is the same or less.) From the rewards given in a
round, player $1$ can tell which of the $k_j$ was active (because $r_1
\neq r_2$). Now, consider the following learning strategy for player $1$:
ignore the rounds in which $k_2$ was active, and use the same learning strategy
as we did for {\em get-close-to-the-target} in the proof of
Theorem~\ref{th:get}, as if $k_1$ was the $k$ of that game. That is,
always play the action in the middle of the remaining interval for
$k_1$, setting $a_1 = \lfloor \frac{k_1^{min}+k_1^{max}}{2}
\rfloor$. The only difference is that we do not update our
stage-game strategy until we {\em lose or draw} a round {\em where
$k_1$ is active}. This is so that we can apply Lemma~\ref{le}: such a
change in strategy will be the end of an epoch. By similar reasoning
as in Theorem~\ref{th:get}, we will know the value of $k_1$ after at
most $\lceil \log(n) \rceil$ epochs (after which there is one more
epoch where we play the maximin strategy $k_1$ and player $2$ can have
no advantage). We now show that the required preconditions of
Lemma~\ref{le} are satisfied. First, if a stage-game strategy for
player $2$ has no chance of changing the epoch, that means that with
that stage-game strategy, player $2$ has no chance of winning or
drawing if $k_1$ is active; it follows that player $2$ can get at most
$p_2r_2 \leq \frac{p_1r_1}{2}$ with this stage-game strategy, and
thus has no advantage. Second, if a stage-game strategy for player
$2$ causes the change with probability $p$, the expected utility of
that stage-game strategy for player $2$ can be at most $pr_1 +
p_2r_2 \leq pr_1 + \frac{p_1r_1}{2}$, so that the expected loss
$\lambda$ in the round to player $1$ is at most $pr_1$. Thus we can
set all the $c_i$ to $r_1$ (apart from the $\lceil \log(n) \rceil+1$th
one which we can set to $0$, because in the corresponding epoch we
will be playing the maximin strategy), and we can conclude by
Lemma~\ref{le} that the game family is BL-WoLF-learnable with loss
$\lceil \log(n) \rceil r_1$.
\end{proof}

We now give an example of a game family with deterministic payoffs
where guaranteed BL-WoLF learning is impossible because the opponent
might be lucky enough to keep winning without revealing any of the
structure of the game.

\begin{game_description}
For given $m>0$ and $n$, the game family {\em
generalized-matching-pennies-with-duds} is defined as follows. Players
1 and 2 both have action space $A=\{1,2,\ldots,m+n\}$. The outcome
function is defined by a subset $D \subseteq A$, with $|D|=n$, of
duds.  Given the actions by the players, the outcome of the game is as
follows (the winner gets $1$, the loser $0$): if one player plays a
dud and the other does not, the latter wins.  Otherwise, if both
players play the same action, player $2$ wins; and if they play
different actions, player $1$ wins.  {\em Player $1$ initially does
not know:} the subset $D$. (We observe that for $m=2$ and $n=0$, we have the classic
matching-pennies game.)
\end{game_description}

{\em Generalized-matching-pennies-with-duds} is not guaranteed
BL-WoLF-learnable, because for any learning strategy for player $1$, it is
possible that player $2$ will happen to keep picking the same action
as player $1$ in every round. In this case, player $1$ accumulates a
huge loss without learning anything at all about the subset
$B$. Nevertheless, {\em generalized-matching-pennies-with-duds} is
BL-WoLF-learnable, as the next theorem shows.

\begin{thm}
The game family {\em generalized-matching-pennies-with-duds} is
BL-WoLF-learnable with loss $n$.
\end{thm}
\vspace{-.12in}
\begin{proof}
We first observe that player $1$ is guaranteed to win at least $\frac{m-1}{m}$
of the time when randomizing uniformly over all nonduds; this is in
fact the maximin strategy. Now consider the following learning strategy for
player $1$: in every round, randomize uniformly over all the actions
besides the ones player $1$ knows to be duds. We will again use
Lemma~\ref{le}. An epoch here ends when player $1$ can classify
another action as a dud; thus, there can be at most $n+1$ epochs, and
in the last epoch player $1$ is playing the maximin strategy and
player $2$ can have no advantage.  We now show that the required
preconditions of Lemma~\ref{le} are satisfied. First, in any epoch but
the last, player $1$ plays duds with some nonzero probability; and if
player $2$ plays a nondud when player $1$ plays a dud, player $1$ will
realize that it was a dud and the epoch will end. Thus, if player $2$
plays a nondud with nonzero probability, the epoch will end with some
probability. On the other hand, if player $2$ always plays duds,
player $2$ will win only if player $1$ happens to play the same dud,
which will happen with probability at most $\frac{1}{q}$ where $q$ is
the number of actions player $1$ is randomizing over. Because $q>m$,
this means player $2$ wins with probability less than $\frac{1}{m}$,
and thus gets no advantage from this. So the first precondition is
satisfied. Second, if in a given epoch where player $1$ is randomizing
over $q$ actions (the $m$ nonduds plus $q-m$ duds), player $2$ plays a
stage-game strategy that plays a nondud with probability $p$, this
will end the epoch with probability at least $p \frac{q-m}{q}$. Also,
the probability that player $2$ wins is at most $p \frac{q-m}{q} +
\frac{1}{q} < p \frac{q-m}{q} + \frac{1}{m}$, so that the expected
loss $\lambda$ in the round to player $1$ is at most $p
\frac{q-m}{q}$. Thus we can set all the $c_i$ to $1$ (apart from the
$n+1$th one which we can set to $0$, because in the corresponding
epoch we will be playing the maximin strategy), and we can conclude by
Lemma~\ref{le} that the game family is BL-WoLF-learnable with loss
$n$.
\end{proof}

\vspace{-.06in}
\section{Nonguaranteed approximate BL-WoLF-learnability}
\label{se:na}

\begin{defin}
A parameterized family of games is {\em approximately
BL-WoLF-learnable} with loss $l$ and precision $\epsilon$ if there
exists a learning strategy for player $1$ such that, for any game in the
family, against any opponent, and for any integer $N$, player $1$'s
expected loss over the first $N$ rounds is at most $l+N\epsilon$.
\end{defin}

We now show that approximate BL-WoLF-learnability is indeed a weaker
notion than guaranteed approximate BL-WoLF-learnability.

\begin{thm}
If a parameterized family of games is guaranteed approximately
BL-WoLF-learnable with loss $l$ and precision $\epsilon$, it is also
approximately BL-WoLF-learnable with loss $l$ and precision
$\epsilon$.
\end{thm}
\vspace{-.12in}
\begin{proof}
Given the learning strategy $\sigma$ that will allow player $1$ to learn enough
about the game to construct an $\epsilon$-approximate maximin strategy
with loss at most $l$, consider the learning strategy $\sigma'$ which plays
$\sigma$ until the $\epsilon$-approximate maximin strategy has been
learned, and plays the $\epsilon$-approximate maximin strategy forever
after that. Then, after $N$ rounds, if we are given that no
$\epsilon$-approximate maximin strategy has been learned yet, the loss
must be less than $l$. Given that an $\epsilon$-approximate maximin
strategy was learned after $i \leq N$ rounds, the loss up to and
including the $i$th round must have been less than $l$, and the
expected loss after round $i$ is at most $(N-i)\epsilon$ (because an
$\epsilon$-approximate maximin strategy was played in every round
after this). It follows that the expected loss is at most
$l+N\epsilon$.
\end{proof}

A version of Lemma~\ref{le} for approximate learning that takes
advantage of the fact that we are allowed to lose $\epsilon$ per round
is straightforward to prove. We will not give it or any examples of
its application here, because of space constraint.

\section{Conclusions and future research}

We presented a general framework for characterizing the cost of
learning to play an unknown repeated zero-sum game.  In our model, the
game falls within some family that the learner knows, and subject to
that, the game is adversarially chosen.  In playing the game, the
learner faces an opponent who knows the game and the learner's
learning strategy.  The opponent tries to give the learner high losses
while revealing little about the game.  Conversely, the learner tries
to either not accrue losses, or to quickly learn about the game so as
to be able to avoid future losses (this is consistent with the {\em
Win or Learn Fast (WoLF)} principle).  Our framework allows for both
probabilistic and approximate learning.  

In short, our framework allows one to measure the worst-case cost of
lack of knowledge in repeated zero-sum games.  This cost can then be
used to compare the learnability of different families of zero-sum
games.

We first introduced the notion of {\em guaranteed
BL-WoLF-learnability}, where a smart learner is guaranteed to have
learned enough to play a maximin strategy after losing a given amount
(against any opponent).  We also introduced the notion of {\em
guaranteed approximate BL-WoLF-learnability}, where a smart learner
is guaranteed to have learned enough to play an $\epsilon$-approximate
maximin strategy after losing a given amount (against any opponent).

We then introduced the notion of {\em BL-WoLF-learnability} where a
smart learner will, {\em in expectation}, lose at most a given amount
that does not depend on the number of rounds (against any opponent).
We also introduced the notion of {\em approximate
BL-WoLF-learnability}, where a smart learner will, {\em in
expectation}, lose at most a given amount that does not depend on the
number of rounds, plus $\epsilon$ times the number of rounds (against
any opponent).  We showed, as one would expect, that if a game family
is guaranteed (approximately) BL-WoLF-learnable, then it is also
(approximately) BL-WoLF-learnable in the weaker sense.

We presented guaranteed BL-WoLF-learnability results for families of
games with deterministic payoffs (namely, the families {\em
get-close-to-the-target} and {\em
generalized-rock-paper-scissors-with-duds}). We also showed that even
families of games with stochastic payoffs can be guaranteed
BL-WoLF-learnabile (for example, the {\em
random-orientation-generalized-rock-paper-scissors-with-duds} game
family).  We also demonstrated that these families are guaranteed
approximate BL-WoLF-learnable with lower cost.

We then demonstrated families of games that are not guaranteed
BL-WoLF-learnable---some of which have stochastic payoffs (for
example, the {\em get-close to-one-of-two-targets} family) and some of
which have deterministic payoffs (for example, the {\em
generalized-matching-pennies-with-duds} family).  We showed that those
families, nevertheless, are BL-WoLF-learnable.  To prove these
results, we used a key lemma which we derived.

Future research includes giving general characterizations of families
of zero-sum games that are BL-WoLF learnable with some given cost
(for each of our four definitions of BL-WoLF learnability)---as well
as characterizations of families that are not.  Future work also
includes applying these techniques to real-world zero-sum games.

{\scriptsize
\bibliography{/afs/cs.cmu.edu/user/amem/refs/dairefs.bib}
\bibliographystyle{mlapa} }
\end{document}